\documentclass[aps,prd,showpacs,preprintnumbers,nofootinbib,twocolumn]{revtex4}
\usepackage{dcolumn}
\usepackage{lipsum}
\usepackage{graphicx,epsfig,float}
\usepackage{psfrag}
\usepackage{bm}
\usepackage{epstopdf}
\usepackage{latexsym}
\usepackage{multirow}
\usepackage{amsmath,amssymb,amsfonts}
\usepackage{enumerate}
\usepackage{hyperref}
\usepackage{multirow}
\usepackage{dcolumn}
\usepackage{array}

\def\beq {\begin{equation}}
\def\eeq {\end{equation}}
\def\bea {\begin{eqnarray}}
\def\eea {\end{eqnarray}}
\def\br{\begin{eqnarray}}
\def\er{\end{eqnarray}}
\def\nn {\nonumber}
\def\bc {\begin{center}}
\def\ec {\end{center}}
\def\bi {\begin{itemize}}
\def\ei {\end{itemize}}
\newcommand{\eel}[1] {\label{#1}\end{equation}}

\newcolumntype{L}[1]{>{\raggedright\arraybackslash}p{#1}}
\newcolumntype{C}[1]{>{\centering\arraybackslash}p{#1}}
\newcolumntype{R}[1]{>{\raggedleft\arraybackslash}p{#1}}

\def\a  {\alpha}

\def\d  {\delta}

\def\e  {\epsilon}

\def\k  {\kappa}
\def\m  {\mu}
\def\n  {\nu}
\def\O  {\Omega}

\def\th {\theta}

\def\t  {\tau}

\def\pa {\partial}

\def\sq {\sqrt}

\def\thin{\thinspace}

\def\l{\left}
\def\r{\right}
\def\dis{\displaystyle}

\def\til{\tilde}

\begin{document}

\title{Quantum entanglement and Hawking temperature}

\author{S.\ Santhosh Kumar } \email[email: ]{santhu@iisertvm.ac.in}
\author{S.\ Shankaranarayanan} \email[email: ]{shanki@iisertvm.ac.in}

\affiliation{School of Physics, Indian Institute of Science Education and Research  Thiruvananthapuram (IISER-TVM), Kerala- 695 016, India}

%
\begin{abstract}
	The thermodynamic entropy of an isolated system is given by
its von Neumann entropy. Over the last few years, there is an 
intense activity to understand thermodynamic entropy 
from the principles of quantum mechanics. More specifically,
is there a relation between the (von Neumann) entropy of entanglement 
between a system and some (separate) environment is related to 
the thermodynamic entropy? It is difficult to obtain the relation 
for many body systems, hence, most of the work in the literature 
has focused on small number systems. In this work, we consider 
black-holes --- that are simple yet macroscopic systems --- 
and show that a direct connection could not be made
between the entropy of entanglement and the Hawking temperature. In
this work, within the adiabatic approximation, we explicitly show that the Hawking temperature is indeed
given by the rate of change of the entropy of entanglement across a
black hole's horizon with regard to the system energy. This is yet another 
numerical  evidence to understand the  key features of black hole thermodynamics  
from the viewpoint of quantum information theory.
\end{abstract}
\maketitle
%



\section{Introduction}
Equilibrium statistical mechanics allows a successful description 
of the thermodynamic properties of matter \cite{birkhoff1931-PNAS,Neumann1932-PNAS,boltzmann1995,L&L-5}. 
 More importantly, it relates entropy, a phenomenological quantity in thermodynamics, to the volume 
of a certain region in phase space \cite{Wehrl1978-RMP}. 
The laws of thermodynamics are also equally applicable to  quantum mechanical 
systems.  A lot of progress has been made recently in 
studying the cold trap atoms that are largely isolated from surroundings 
\cite{weiss2006-nature,gross2008-nature,smith2013-njp,yukalov2007-LPL}.  Furthermore, the availability of Feshbach 
resonances is shown to be useful to control the strength of interactions, 
to realize  strongly correlated systems, and to drive these systems between 
different quantum phases in controlled manner \cite{Osterloh2002,wu2004,rey2010-njp,santos2010-njp}.  These experiments have raised the possibility of understanding the emergence of 
thermodynamics from principles of quantum mechanics. The fundamental questions 
that one hopes to answer from these investigations are:   How the macroscopic 
laws of thermodynamics emerge from the reversible quantum dynamics? How to 
understand the thermalization of a closed quantum systems? 
What are the relations between information, thermodynamics and quantum mechanics
\cite{2006-Lloyd-NPhys,2008-Brandao,horodecki-2008,popescu97,vedral98,plenio98} ? While answer to these questions, for many body system is out of sight, some 
important progress has been made by considering simple lattice systems  (See, for 
instance, Refs. \cite{1994-srednicki,rigol2008-nature,2012-srednicki,rahul2015-ARCMP}). In this work,  in an attempt to address some of the above questions, 
our focus is on another simple, yet, macroscopic system --- black-holes.

It has long been conjectured that a black hole's thermodynamic entropy is given by its entropy of entanglement 
across the horizon~\cite{bombelli86,srednicki93,eisert2005,shanki2006,shanki-review,solodukhin2011,shanki2013}.
However, this has never been directly related to the Hawking temperature~\cite{hawking75}. Here we
show that:
\begin{enumerate}[(i)]
 \item Hawking temperature is given by the rate of change of the entropy of 
entanglement across a black hole's horizon with regard to the system energy.  
\item The information lost across the horizon is related to black hole entropy 
and laws of black hole mechanics emerge from entanglement across 
the horizon. 
\end{enumerate}

The model we consider is complementary to other models that 
investigate the emergence of  thermodynamics \cite{2006-Lloyd-NPhys,2008-Brandao,horodecki-2008,popescu97,vedral98,plenio98}: 
First, we evaluate the entanglement entropy for a relativistic 
free scalar fields propagating in the black-hole background while the simple lattice models 
that were considered are non-relativistic. Second,  quantum entanglement can be
unambiguously quantified only for bipartite systems~\cite{horodecki2009,eisert2010}. 
While the bipartite system is an approximation for applications to many body systems, 
here, the event horizon provides a natural boundary.

Evaluation of the entanglement of a relativistic free scalar field, as always, is the simplest 
model. However, even for free fields it is difficult to obtain the entanglement entropy. 
The free fields are Gaussian and these states are entirely characterized by 
the covariance matrix. It is generally difficult to handle covariance matrices in an infinite 
dimensional Hilbert space \cite{eisert2010}. There are two ways to calculate entanglement 
entropy in the literature. One approach is to use the replica trick  which rests on evaluating 
the partition function on an n-fold cover of the background geometry where a cut is 
introduced throughout the exterior of the entangling surface \cite{eisert2010,cardy2004}. 
Second is a {\it direct approach}, where the Hamiltonian of the field is discretized 
and the reduced density matrix is evaluated in the real space.  We adopt this approach 
as entanglement entropy may have more
symmetries than the Lagrangian of the system~\cite{krishnand2014}.

To remove the spurious effects due to the coordinate singularity at
the horizon\footnote{In Schwarzschild coordinate, $r >  2M$ need to be bipartited \cite{1998-Mukohyama}.}, we consider Lema\^itre coordinate which is explicitly
time-dependent\cite{L&L-2}. One of the features that we exploit in our
computation is that for a fixed Lema\^itre time coordinate,
Hamiltonian of the scalar field in Schwarzschild space-time reduces to
the scalar field Hamiltonian in flat space-time \cite{shanki-review}.

The procedure we adopt is the following: 
\begin{enumerate}[(i)]
\item We perturbatively evolve the Hamiltonian about the fixed Lema\^itre
time.  
\item We obtain the entanglement entropy at different times. We show
that at all times, the entanglement entropy satisfies the area law i. e. 
$S(\epsilon) = C(\epsilon) A$ where $S(\epsilon)$ is the entanglement entropy evaluated at a given Lema\^itre time $(\epsilon)$, $C(\epsilon)$ is the proportionality constant that depends on $\epsilon$, and $A$ is the area of  black hole horizon. In other words,
the value of the entropy is different at different times. 
\item We calculate the change in entropy as function of $\epsilon$, 
i. e., $\Delta S/\Delta \epsilon$. Similarly we calculate change in energy $E(\e)$, i.e., 
$\Delta E/\Delta \epsilon$. 
\end{enumerate}
For several black-hole metrics, we explicitly show that ratio of the rate of change of energy and the rate of change of entropy is identical to the Hawking temperature.

The outline of the paper is as follows: In Sec. (\ref{sec.1}), we set up our model Hamiltonian to obtain the entanglement entropy in ($D+2$)-dimensional space time. Also, we define {\it entanglement temperature}, which had the same structure from the statistical mechanics, that is, ratio of change in total energy to change in entanglement entropy. In Sec. (\ref{sec.2}), we numerically show that for different black hole space times, the divergent free {\it entanglement temperature} matches approximately with the Hawking temperature obtained from general theory of relativity and its Lovelock generalization. This provides a strong evidence towards the interpretation of entanglement entropy as the Bekenstein-Hawking entropy. Finally in Sec. (\ref{sec.3}), we conclude with a discussion to connect our analysis with the eigenstate thermalization hypothesis for the closed quantum systems \cite{2012-srednicki}.  

Throughout this work, the metric signature we adopt is
$(+,-,-,-)$ and set $\hbar=k_{B} =c=1$.
\section{Model and Setup}
\label{sec.1} 
\subsection{Motivation}
Before we go on to evaluating entanglement entropy (EE) of a quantum scalar field propagating in black-hole background, 
we briefly discuss the motivation for the studying entanglement entropy of a scalar field. Consider the Einstein-Hilbert 
action with a positive cosmological constant ($|\Lambda|$):
\beq
\label{eq:EHAction}
\mathcal S_{_{EH}} (\bar{g}) = M_{_{\rm Pl}}^2 \int d^4x \sqrt{-\bar{g}}
\left[\bar{R} - 2 |\Lambda|\right] \, .  \eeq
{where $\bar R$ is the Ricci scalar and $M_{_{\rm Pl}}$ is the Plank mass.  }
Perturbing the above action w.r.t. the metric $\bar{g}_{\mu\nu} = g_{\mu\nu} + h_{\mu\nu}$, 
the action up to second order becomes~\cite{shanki-review}:
\bea
\mathcal S_{_{EH}}(g, h) &=& - \frac{M_{_{\rm Pl}}^2}{2} \int \, d^4x \sqrt{|g|}\,
\left[\nabla_{\alpha} {h}_{\mu\nu} \nabla^{\alpha} {h}^{\mu\nu}\r.\nn\\
&& \l. \dis+ |\Lambda| h_{\mu\nu} h^{\mu\nu}\right]\!\! \, .
\eea
The above action corresponds to  massive ($\Lambda$) spin-2 
field ($h_{\mu\nu}$) propagating in the background metric $g_{\mu\nu}$. 
Rewriting, 
$h_{\mu\nu} = M_{_{\rm Pl}}^{-1} \epsilon_{\mu\nu} \Phi(x^{\mu})$
[where $\epsilon_{\mu\nu}$ is the constant polarization tensor], the
above action can be written as
\beq
\mathcal S_{_{EH}} (g, h) = - \frac 1 2 \int \, d^4x \sqrt{|g|}\,
\left[\pa_{\alpha} \Phi \pa^{\alpha} \Phi
+ |\Lambda| \Phi^2 \right] \, .
\eeq
which is the action for the massive scalar field propagating in the
background metric $g_{\mu\nu}$. In this work, we consider massless ($\Lambda = 0$ corresponding 
to asymptotically flat space-time) scalar field propagating in $(D + 2)-$dimensional spherically symmetric space-time.

\subsection{Model}

The canonical action for the massless, real scalar field
$\Phi(x^{\mu})$ propagating in $(D + 2)-$dimensional space-time is

\beq
 \label{equ21}
 \mathcal{S}=\dis\frac{1}{2}\int d^{D+2}{\bf x}\thin \sqrt{-g}\thin g^{\m \n}\partial_{\m}\Phi({\bf x})\thin\partial_{\n}\Phi({\bf x})
 \eeq
where $g_{\m\n} $ is the spherically symmetric Lema\^{\i}tre line-element \cite{L&L-2}:
 \beq
 \label{equ22} 
 ds^2=
 d\t^2-\l(1-f[r(\t,\xi)]\r)\,d\xi^2-r^2(\t,\xi)d\O^2_D
 \eeq
where $\t,\xi$ are the time and radial components in Lema\^itre 
coordinates, respectively, $r$ is the radial distance in Schwarzschild coordinate and $d\O_D$ 
is the $D-$dimensional angular line-element. In order for the line-element (\ref{equ22}) to describe a black hole,
the space-time must contain a singularity (say at $r = 0$) and have
horizons. We assume that the  asymptotically flat space-time 
contains one non-degenerate event-horizon at $r_h$. The specific form of $f(r)$ corresponds to different space-time.

Lema\^itre coordinate system has the following interesting properties: 
\begin{enumerate}[(i)]
 \item The coordinate $\tau$ is time-like all across $0< r < \infty$, similarly $\xi$ is space-like all across $0< r < \infty$.
 \item Lema\^itre coordinate system does not have coordinate singularity at the horizon. 
\item This coordinate system is time-dependent. The test particles at rest relative to the  reference system are particles moving freely in the given field \cite{L&L-2}.
 \item Scalar field propagating in this coordinate system is explicitly time-dependent. 
 \end{enumerate}

The spherical symmetry of the line-element (\ref{equ22}) allows us to decompose the normal modes of the scalar field as:
 \beq
\label{equ23}
\Phi({\bf x})=\dis\sum_{l,m_i}\Phi_{lm_i}(\t,\xi) \,Z_{_{lm_i}}(\th,\phi_i),
\eeq
\noindent where $i \in \{1,2,\ldots D-1\}$ and $Z_{_{lm_i}}$'s are the
real hyper-spherical harmonics. We define the following dimensionless
parameters:
$\til{r}=r/r_h, \thin \til{\xi}=\xi/r_h,\thin \til{\t}=\t/r_h, \thin\til{\Phi}_{lm}=r_h \, \Phi_{lm}. $
By the substitution of the orthogonal properties of $Z_{_{lm_i}}$, the canonical massless scalar field action becomes, 
\br
\label{equ26}
\mathcal{S}&=&\dis\frac{1}{2}\sum_{_{l,m_i}}\int d\til{\t}\thin d\til{\xi} \thin \til{r}^D \l[\sq{1-f[\til{r}]}\,\, (\partial_{\tilde{\t} }\til{\Phi}_{lm_i})^2
-\frac{1}{\sq{1-f[\til{r}]}}\r.\nn\\
&&\l. \dis \times (\partial_{\tilde{\xi }}\til{\Phi}_{lm_i})^2-\sqrt{1-f[\til{r}]}\,\,\frac{ l(l+D-1)}{\til{r}^2}\til{\Phi}_{lm_i}^2\r]
\er
The above action contains non-linear time-dependent terms through $f(\til r)$. Hence, the Hamiltonian 
obtained from the above action will have non-linear time-dependence. While the full non-linear 
time-dependence is necessary to understand the small size black-holes, for large size black-holes, 
it is sufficient to linearize the above action by fixing the time-slice and performing the  
following  infinitesimal transformation  about a particular Lema\^itre time $\til \t$ \cite{toms}. More specifically,
\begin{subequations}
\br
	\til{\t}\rightarrow \til{\t}'=\til{\t}+\e, \qquad \til{\xi}\rightarrow \til{\xi}'=\til{\xi},\\	\til{r}(\til{\t}',\til{\xi}')=\til{r}(\til{\t}+\e,\til{\xi}),\\
	\til{\Phi}_{lm_i}(\til{\t},\til{\xi})\rightarrow \til{\Phi}'_{lm_i}(\til{\t}',\til{\xi'})=\til{\Phi}_{lm_i}(\til{\t},\til{\xi})
	\label{equ27}
\er
\end{subequations}
where $\e$ is the infinitesimal Lema\^itre time. The functional expansion of $f(\til r)$ about $\e$ and  the following relation between the  Lema\^itre coordinates \cite{L&L-2},
\beq
\til\xi - \til\t=\int\frac{d\til r}{\sq{1- f[\til r(\til\t, \til\xi)]}},
\eeq 
allow us to perform the perturbative expansion in the above action. 

After doing the Legendre transformation, the Hamiltonian up to second order in $\e$ is 
\beq
\label{Hamilt_1}
H (\e)\simeq H_{_0}+ \e V_{_1}+\e^2 V_{_2} 
\eeq
 where $ H_{_0}$ is the unperturbed scalar field Hamiltonian in the flat space-time, $V_{_1} \,\mbox{and}\,V_{_2}$ are 
 the perturbed parts of the Hamiltonian (for details, see Appendix \ref{app1}). Physically, the above infinitesimal 
 transformations (\ref{equ27}) correspond to perturbatively expanding the scalar field about a particular 
 Lema\^itre time.

\subsection{Important observations}
The Hamiltonian in Eq. (\ref{Hamilt_1}) is key equation regarding which we would like to
stress the following points: First, in the limit of $\epsilon \to 0$,
the Hamiltonian reduces to that of a free scalar field propagating in
flat space-time \cite{shanki-review}.  In other words, the zeroth order
Hamiltonian is identical for all the space-times.  Higher order
$\epsilon$ terms contain information about the global space-time
structure and, more importantly, the horizon properties. 

Second, the Lema\^itre coordinate is intrinsically time-dependent; the $\epsilon$
expansion of the Hamiltonian corresponds to the perturbation about the
Lema\^itre time. Here, we assume that the Hamiltonian $H$ undergoes adiabatic 
evolution and the ground state $\Psi_{GS}$ is the instantaneous ground state at all Lema\^itre times. 
This assumption is valid for large black-holes as Hawking evaporation is not significant. Also, since the line-element 
is time-asymmetric, the vacuum state is Unruh vacuum. Evaluation of the entanglement entropy  for
different values of $\epsilon$ corresponds to different values of
Lema\^itre time. As we will show explicitly in the next section, entanglement entropy at a given $\epsilon$ 
satisfies the area law  [$S(\epsilon) \propto A$] and the proportionality constant depends on $\epsilon$ i. e. $S(\epsilon) = C(\epsilon) A$. 

Third, it is not possible to obtain a closed form
analytic expression for the density matrix (tracing out the quantum
degrees of freedom associated with the scalar field inside a spherical
region of radius $r_h$) and hence, we need to resort to numerical
methods. In order to do that we take a spatially uniform radial grid,
$\{  r_j\}$, with $b =  r_{j + 1} - r_j$. We discretize the Hamiltonian
$H$ in Eq.(\ref{Hamilt_1}).
The procedure to obtain the entanglement entropy for different
$\epsilon$ is similar to the one discussed in
Refs.~\cite{srednicki93,shanki-review}. In this work, we assume that
the quantum state corresponding to the discretized Hamiltonian
is the ground state
with wave-function $\Psi_{GS}(x_1,\ldots,x_n;y_1,\ldots,y_{N- n})$. The
reduced density matrix $\rho(\vec y,{\vec y\,}')$ is obtained by tracing
over the first $n$ of the $N$ oscillators
\beq
\rho(\vec y,{\vec y\,}') =\dis
\int \dis(\prod_{i =1}^{n} dx_i)\,
\Psi_{GS}(x_1,..,x_n;\vec y\,)
\Psi^{*}_{GS}(x_1,..,x_n;{\vec y\,}')
\eeq

Fourth, in this work, we use von Neumann entropy 
\beq
\label{renyi}
 S(\epsilon) = -\mbox{Tr}\l(\rho \;\log \rho\r)
 \eeq
%
as the measure of entanglement. 
In analogy with microcanonical ensemble picture of equilibrium
statistical mechanics, evaluation of the Hamiltonian $H$ at different infinitesimal Lema\^itre time
$\epsilon$, corresponds to setting the system at different internal
energies. In analogy we define {\it entanglement temperature}
\cite{sakaguchi89}:
\beq
\label{temp_1}
\frac{1}{T_{EE}} = \frac{\dis \Delta S(\epsilon)
}{\dis\Delta E(\epsilon)}= \frac{\mbox{Slope of
    EE}(\dis\Delta S/\Delta\e)}{\mbox{Slope of
    energy}(\dis\Delta E/\Delta\e)} \eeq
%

The above definition is consistent with the statistical mechanical definition of temperature. In 
statistical mechanics, temperature is obtained by evaluating change in the entropy and energy 
w.r.t. thermodynamic quantities. In our case, entanglement entropy and energy depend 
on the Lema\^itre time, we have evaluated the change in the entanglement entropy and energy 
w.r.t. $\epsilon$.  In other words, we calculate the change in the ground state energy
(entanglement entropy) for different values of $\epsilon$ and find the
ratio of the change in the ground state energy and change in the
EE. As we will show in the next section, EE and energy goes linearly 
with $\epsilon$ and hence, the temperature does not depend on $\epsilon$.
While the EE and the energy
diverge, their ratio is a non-divergent quantity.  To understand this,
let us do a dimensional analysis
\begin{subequations}
\bea [E_{D+1}]\propto N^{D+1}\propto \dis \l(\frac{L}{b}\r)^{D+1},~~
     [S]\propto \dis \frac{A_D}{b^D} \\ \Rightarrow \thinspace
     [\til T_{EE}]\propto \dis \frac{[E_{D+1}]}{[S]}\propto N
     \dis\frac{L^D}{A_D} \\ \mbox{i.e,} ~~~ \dis
     \frac{[\til T_{EE}]}{N}=[T_{EE}] \propto(N/n)^D\Rightarrow
     \mbox{finite}~~~~~~
\label{temper_1}
\eea
\end{subequations}
where $A_D$ is the $D+1$ dimensional hyper- surface area. In the thermodynamic limit, by
setting $L$ finite with $N \to \infty$ and $b \to 0$, $T_{EE}$ in
Eq.~(\ref{temper_1}) is finite and independent of $\e$.
 
 For large $N$, we show that, in the natural units, the above
 calculated temperature is identical to Hawking temperature for the
 corresponding black-hole \cite{hawking75}:
\beq
\label{eq:HawkingTemp}
T_{BH} = \l. \frac{\k}{2\pi}=\frac{1}{4\pi}\frac{d f( r)}{d
  r}\r|_{_{ r= r_h} } \eeq
Fifth, it is important to note the above {\it entanglement
  temperature} is non-zero only for $f(r) \neq 1$. In the case of flat
space-time, our analysis shows that the {\it entanglement temperature} vanishes, and we obtain $T_{EE}$ numerically for different black hole space-times.
\section{Results and Discussions}
\label{sec.2}
The Hamiltonian $H$ in Eq. (\ref{Hamilt_1}) is mapped to a system of $N$ coupled  time independent harmonic oscillators (HO) 
with non-periodic boundary conditions. The interaction matrix elements of the Hamiltonian can be found in Ref \cite{dropbox}. 
The total internal energy (E) and the entanglement entropy ($S$) for the ground state of the HO's is computed numerically 
as a function of $\e$  by using central difference scheme (see Appendix \ref{app2}). 
 All the computations are done using MATLAB R$2012$a for the lattice size  $N = 600$, $ 10 \leq n \leq 500$ 
 with a minimum accuracy of $10^{-8}$ and a maximum accuracy of $10^{-12}$. 

  In the following subsections, we compute $T_{EE}$ numerically
    for two different black-hole space-times, namely, 4 dimensional Schwarzschild and Reissner-Nordstr\"om black holes  and show that
      they match with Hawking temperature $T_{BH}$.  
       $T_{EE}$ is calculated by taking the average of  {\it entanglement temperature} for each $n$'s by fixing $N$.

\subsection{Schwarzschild (SBH) black holes} 
The 4-dimensional Schwarzschild
black hole space-time ( put $D=2$) in dimensionless units $\til r$ is given by the
line element in Eq.(\ref{equ22}) with $f(\til r)$ is given by:
\beq
f(\til r)=1- \frac{1}{\til r} 
\eeq

\begin{figure}
	\includegraphics[scale=.64]{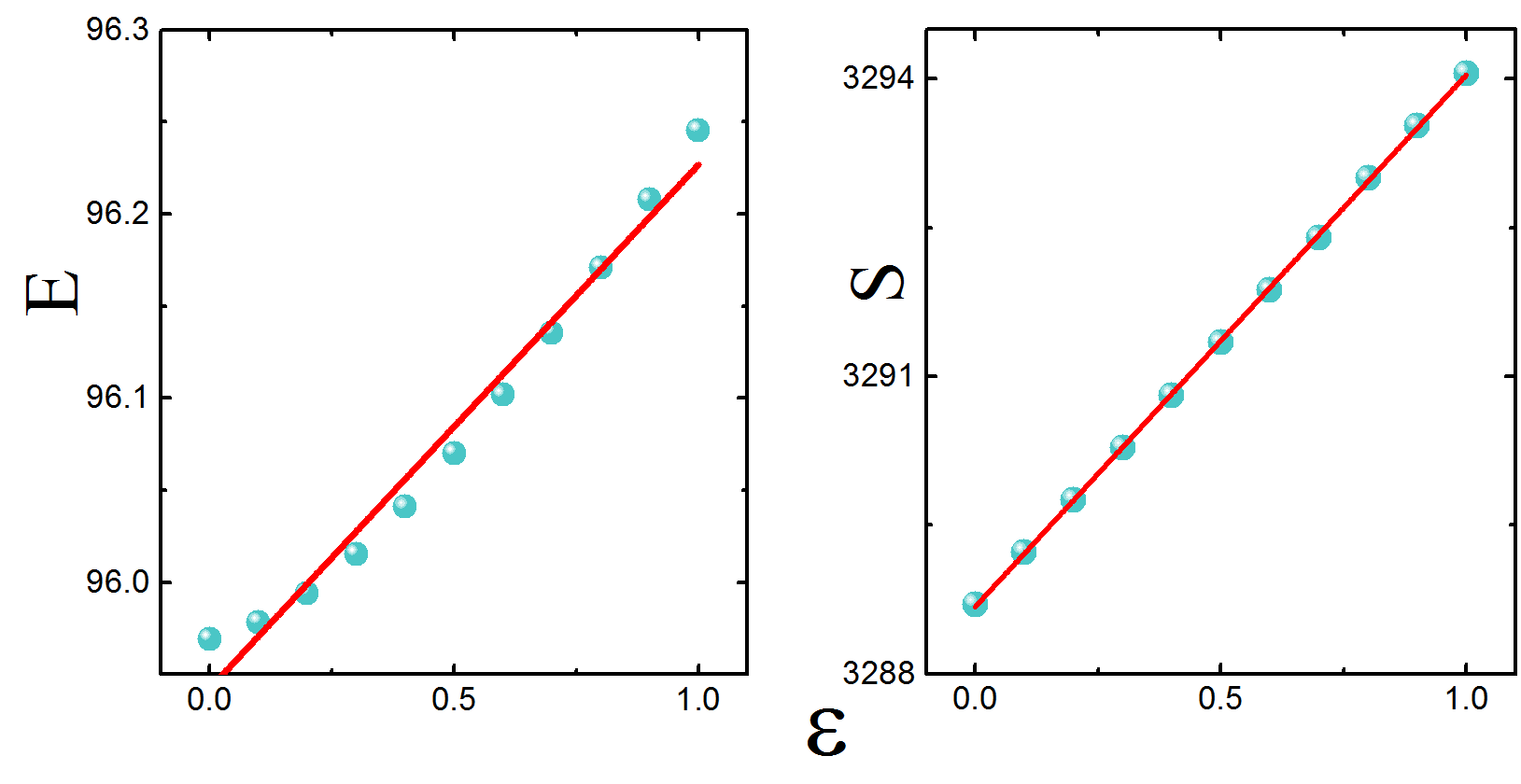}
	\caption{The plots of total energy (left) and EE (right) as a function of $\e$ for the 4d  Schwarzschild black hole. We set $N = 600$ and $n=150$. The cyan coloured dots are the numerical data and the red line is the best linear fit to the data.}
	 \label{fig1}
	\end{figure}

In Fig.(\ref{fig1}), we have plotted total energy (in dimensionless
units) and EE versus $\e$ for 4-dimensional Schwarzschild
space-time.  
Following points are important to note regarding the numerical results:
First for every $\e$, von Neumann entropy scales approximately as
$S \sim (r_h/b)^2$. Second, EE and the total energy increases
with $\e$.  

Using relation (\ref{temp_1}), we evaluate ``entanglement'' temperature
numerically. In dimensionless units, we get
$T_{EE}=0.0793$ which is close to the value of the Hawking temperature
$0.079$.
 However, it is important to
note that for different values of $N$, we obtain approximately the
same value of entropy. The results are tabulated, see
Table(\ref{table1}). 
See  Appendix \ref{app3}, 
for plots of energy and EE for   $n=50,80, 100$ and  $130$. 
	
\subsection{ Reissner-Nordstr\"om (RN) black holes} 
The 4-dimensional  Reissner-Nordstr\"om black
hole is given by the line element in Eq.(\ref{equ22}), where $f(\til
r)$ is 
\beq
\label{equ219}
f(\til r)=1-\frac{2M/r_h}{\til r}+\frac{\l(Q/r_h\r)^2}{\til r^2}
\eeq
 $Q$ is the charge of the black hole. Note that we have  rescaled the radius w.r.t the outer horizon ($ r_h=M+\sqrt{M^2-Q^2}$).
Choosing $q=Q/r_h$, we get 
\beq
f(\til r) = 1-\frac{(1+q^2)}{\til r}+\frac{q^2}{\til r^2}
\eeq
and the black hole temperature in the unit of $r_h$ is 
$T_{BH} = (1-q^2)/4\pi$.
\begin{figure*}
		\includegraphics[scale=.75]{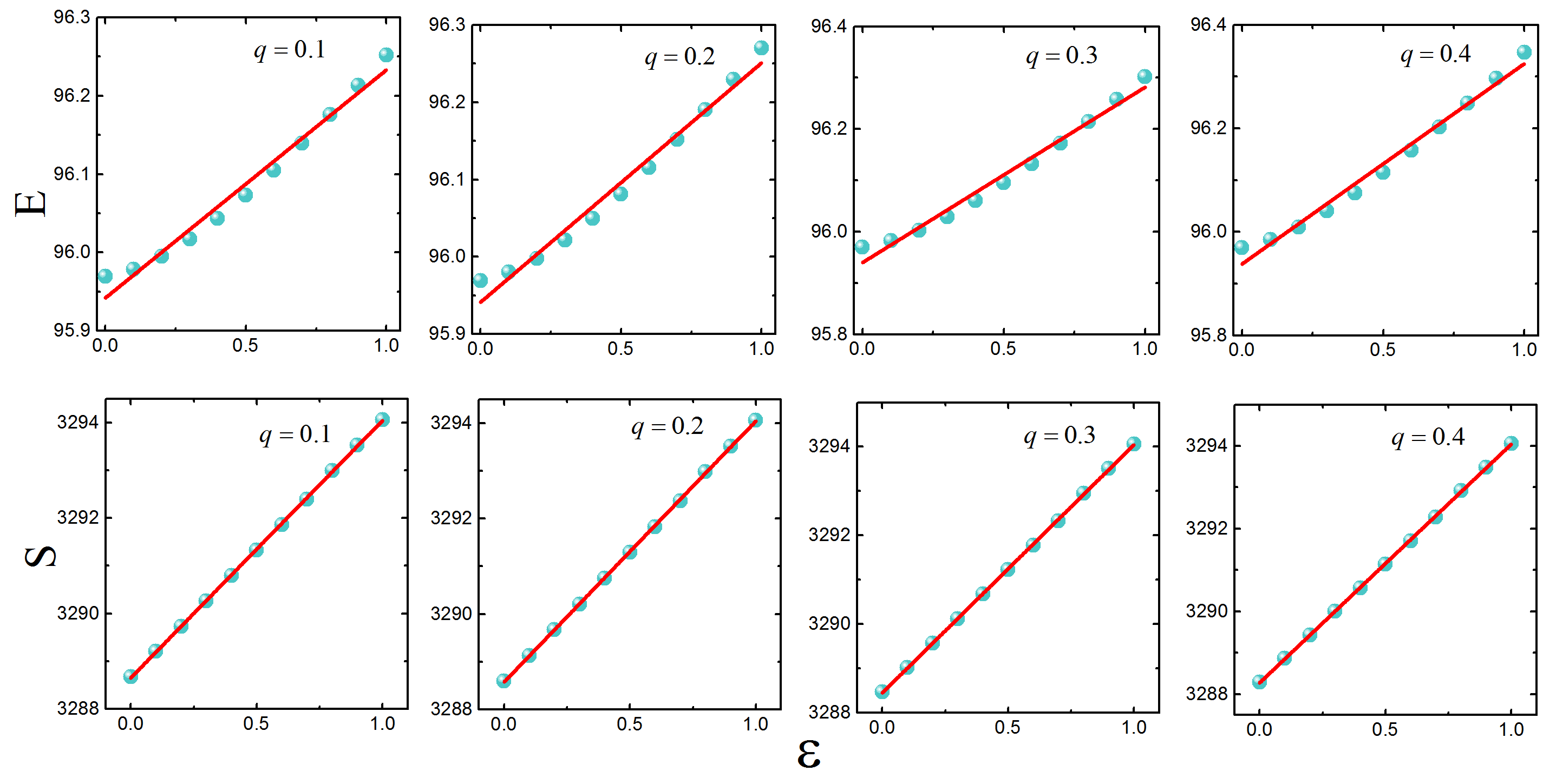}
	\caption{The plots of total energy (top panel) and EE (bottom panel) for as a function of $\e$ for different $q$'s in 4- dimensional Reissner-Nordstr\"om black hole. For these, we set $N = 600$ and $n=150$. The cyan coloured dots are the numerical data and the red line is the best linear fit to the data.}
	 \label{fig3}
	\end{figure*}

Note that we have evaluated the {\it entanglement temperature} by fixing the charge $q$. For  a fixed charge $q$, the first law of black hole mechanics is given by $ d E=\l(\k/2\pi\r) dA$, where $A$ is the area of the black hole horizon. The energy and EE for different $q$ values have the same profile,
which looks exactly like in the previous case and is shown in the 
middle row in Fig. (\ref{fig3}). See Appendix \ref{app3}, for plots for other 
values of $n$. As given in the table (\ref*{table1}), $T_{EE}$
matches with Hawking temperature.

 \begin{table}
\centering
  \begin{tabular}{|C{2cm} |C{1.5cm} |C{1.5cm}|C{1.5cm}|}
  \cline{1-4}
 \bf Black hole space time&   & $\mathbf{T_{BH}}$ & $\mathbf{T_{EE}}$\\ \cline{1-4}
  {\bf 4d-SBH} 
 & &  0.07958& 0.07927\\\cline{1-4} 
  
  	\multirow{3}{*}{{\bf{4d-RN}}} 
& $q=0.1$  &  0.07878 & 0.07836\\\cline{2-4}
  & $q=0.2$ & 0.07639  & 0.07507\\\cline{2-4}
  &$q=0.3$& 0.07242  & 0.07501 \\\cline{2-4}
  & $q=0.4$& 0.06685  &  0.06659\\\cline{2-4}
  	\hline  
  \end{tabular}
 \caption{ The table lists {\it entanglement temperature} and Hawking temperature (measure in units of $r_h$ )  for 
  4-dimensional  Schwarzschild and Reissner-Nordstr\"om black hole space times.}
 \label{table1}
\end{table}
\section{ Conclusions and outlook} 
\label{sec.3}
In this work, we have given another proof that 
4-dimensional black hole entropy can be associated to entropy of entanglement across the horizon
by explicitly deriving entanglement temperature. Entanglement temperature is given by the rate of 
change of the entropy of entanglement across a black hole's horizon with regard
to the system energy. Our new result sheds the light on the interpretation of temperature from 
entanglement as the Hawking temperature, one more step to understand the black hole thermodynamics from the 
quantum information theory platform.

Some of the key features of our analysis are: First, while entanglement and energy diverge 
in the limit of $b \to 0$, the {\it entanglement temperature} is a finite quantity. Second, 
{\it entanglement temperature} vanishes for the flat space-time. While the 
evaluation of the entanglement entropy {\it does not} distinguish between the 
black-hole space-time and flat space-time, entanglement temperature distinguishes the two
space-times.

Our analysis also shows that the entanglement entropy satisfies all
the properties of the black-hole entropy.  First, like the black hole
entropy, the entanglement entropy increases and never
decreases. Second, the entanglement entropy and the temperature
satisfies the first law of black-hole mechanics $dE= T_{EE}
\,dS$. We have shown this explicitly for Schwarzschild black-hole and for Reissner-Norstrom black-hole .  

It is quite remarkable that in higher dimensional space time the R\'enyi entropy provides a convergent alternative to the measure of entanglement  \cite{shanki2013}, however, entanglement temperature will depend on the 
R\'enyi parameter. While a physical understanding of the R\'enyi parameter 
has emerged \cite{baez_renyi_2011}, it is still not clear how to fix the 
R\'enyi parameter from first principles \cite{progress}. 

Our analysis throws some light on the emergent gravity paradigm \cite{Sakharov2000,jacobson95,padmanabhan2010,Verlinde2011} 
where gravity is viewed not as a fundamental force. Here we have shown that the 
information lost across the horizon is related to the black-hole entropy and the 
laws of black-hole mechanics emerge from the entanglement across the horizon. 
Since General Relativity  reduces gravity to an effect of the curvature of the space-time, 
it is thought that the microscopic constituents would be the {\it atoms of the space-time} itself. 
Our analysis shows that entanglement across horizons can be used as building blocks of space-time \cite{VanRaamsdonk2010-GRG,VanRaamsdonk2010-IJMP}.

One of the unsettling questions in theoretical physics is whether due
to Hawking temperature the black-hole has performed a non-unitary
transformation on the state of the system aka information loss
problem. Our analysis here does not address this for two reasons: (i)
Here, we have fixed the radius of the horizon at all times and
evaluated the change in the entropy while to address the information
loss we need to look at changing horizon radius. (ii) Here, we have
used perturbative Hamiltonian, and hence, this analysis  fails as the
black-hole size shrinks to half-its-size \cite{Almheiri2013-JHEP}.  We hope to
report this in future.

While the unitary quantum time-evolution is reversible and retains all
information about the initial state, we have shown that the
restriction of the degrees of freedom outside the event-horizon at all
times leads to temperature analogous to Hawking temperature.  Our
analysis may have relevance to the eigenstate thermalization
hypothesis \cite{1994-srednicki,rigol2008-nature,2012-srednicki,rahul2015-ARCMP}, which we plan to explore.
\section*{ACKNOWLEDGMENTS}
Authors wish to thank A. P. Balachandran,
Charles Bennett, Samuel Braunstein, Saurya Das and Jens Eisert for discussions and
comments. Also, we would like to thank the anonymous referee for the useful comments.  All numerical computations  were done at the fast computing clusters at IISER-TVM. 
The work is supported by Max Planck-India Partner Group on
Gravity and Cosmology. SSK acknowledges the  financial support of the CSIR, Govt. of India through Senior Research Fellowship.  
SS is partially supported by Ramanujan Fellowship of DST, India.  
\begin{widetext}
\begin{appendix}
	\section {Calculation of Scalar field Hamiltonian in Lema\^itre coordinate}
	\label{app1}
	In this Appendix section, we give details of the derivation of the Hamiltonian (H) upto second order in $\e$. 
	Using the orthogonal properties of the real spherical harmonics $Z_{_{lm_i}}$, the scalar field action reduces to, 
	\bea
	\label{equu26}
	\mathcal S&=&\dis\frac{1}{2}\sum_{_{l,m_i}}\int d\til{\t}\thin d\til{\xi} \thin \til{r}^D \l[\sq{1-f[\til{r}]} (\partial_{\tilde{\t} }\til{\Phi}_{lm_i})^2
	-\frac{(\partial_{\tilde{\xi }}\til{\Phi}_{lm_i})^2}{\sq{1-f[\til{r}]}} 
\dis-\sqrt{1-f[\til{r}]}\,\frac{ l(l+D-1)}{\til{r}^2}\til{\Phi}_{lm_i}^2\r]
	\eea
	where $\til{r}=r/r_h, \thin \til{\xi}=\xi/r_h,\thin \til{\t}=\t/r_h, \thin\til{\Phi}_{lm}=r_h \, \Phi_{lm} $ are dimensionless. 
	
	Performing
	the following infinitesimal transformation \cite{toms} in the above  resultant action: 
	\begin{subequations}
	\br
	\til{\t}\rightarrow \til{\t}'=\til{\t}+\e,\,\til{\xi}\rightarrow \til{\xi}'=\til{\xi},\\	
	\til{\Phi}_{lm_i}(\til{\t},\til{\xi})\rightarrow \til{\Phi}'_{lm_i}(\til{\t}',\til{\xi'})=\til{\Phi}_{lm_i}(\til{\t},\til{\xi}),\\
	\til{r}(\til{\t}',\til{\xi}')=\til{r}(\til{\t}+\e,\til{\xi})
	\er
		\end{subequations}
%
%
 The  action in Eq. (\ref{equu26}) becomes,
	\br
	\mathcal S
	&\simeq &\dis\frac{1}{2}\sum_{_{l,m_i}}\int d\til{\t}\thin d\til{\xi} \thin 
	\l(\til{r}+\e h_1+\e^2 h_2/2\r)^D\l[\l(1-f-\e h_1\frac{ \pa f}{\pa \til r}-\frac{\e^2}{2}\l[h_2\frac{ \pa f}{\pa \til r}+h_1^2 \frac{ \pa ^2f}{\pa \til r^2}\r]\r)^{1/2}
	(\partial_{\tilde{\t} }\til{\Phi}_{lm_i})^2\r.\nn\\
	& &\l.-\l(1-f-\e h_1\frac{ \pa f}{\pa \til r}-\frac{\e^2}{2}\l[h_2\frac{ \pa f}{\pa \til r}+h_1^2 \frac{ \pa ^2f}{\pa \til r^2}\r]\r)^{-1/2}
	(\partial_{\tilde{\xi} }\til{\Phi}_{lm_i})^2 
	-\frac{l(l+D-1)}{\l(\til{r}+\e h_1+\e^2 h_2/2\r)^2} \r.\nn\\
	& &\l. \times\l(1-f-\e h_1\frac{ \pa f}{\pa \til r}-\frac{\e^2}{2}\l[h_2\frac{ \pa f}{\pa \til r}+h_1^2 \frac{ \pa ^2f}{\pa \til r^2}\r]\r)^{1/2}\til{\Phi}_{lm_i}^2\r]
	\label{equ28}
	\er
	
	%
	where $h_1=\dis\frac{\partial\til r}{\pa\til \t} ~~\mbox{and}~~ h_2=\dis\frac{\pa^2\til r}{\pa\til \t^2} $. 
	
	Using the  relation between the Lema\^itre coordinates \cite{L&L-2}
	\beq
	\xi - \t=\int\frac{dr}{\sq{1- f[r(\t, \xi)]}}
	\eeq 
	gives the following expression,  
	\beq
	\label{equ214}
	h_1=-\sq{1-f},~~h_2=\frac{-1}{2}\frac{\pa f}{\pa\til r},~~
	\frac{d\til\xi}{d\til r}|_{\mbox{con.}\til\t}=\frac{1}{\sq{1-f}}                 
	\eeq

	The Hamiltonian $(H)$ corresponding to the above Lagrangian is  
	{\small
		\beq
		\label{equ212}
		H\simeq\dis\frac{1}{2}\sum_{_{l,m_i}}\int d\til{\xi}\l[\til{\Pi}_{lm_i}^2+\frac{g_{_1}^D}{g_{_2}} 
		\l( \pa_{\til\xi}\frac{\til\chi_{_{lm_i}}}{g_{_1}^{D/2}\sq{g_{_2}}}\r)^2+\frac{l(l+D-1)}{g_{_1}^2}\til\chi_{_{lm_i}}^2\r]
		\eeq
	}
	where
	\beq
	\label{equ210}
	g_{_1}=\til{r}+\e h_1+\e^2 h_2/2,~~
	g_{_2}=\sq{1-f-\e h_1\frac{ \pa f}{\pa \til r}-\frac{\e^2}{2}\l[h_2\frac{ \pa f}{\pa \til r}+h_1^2 \frac{ \pa ^2f}{\pa \til r^2}\r]} ,~~
	\til{\chi}_{_{lm_i}}= g_{_1}^{D/2}\sq{g_{_2}}\til{\Phi}_{lm_i}
	\eeq
	and $\til{\Pi}_{lm_i}$ is the canonical conjugate momenta corresponding to the
	field $\til{\chi}_{lm_i}$.
	
	Upon quantization, $\til{\Pi}_{lm_i}$ and $\til{\chi}_{lm_i}$ satisfy the usual 
	canonical commutation relation:
	\beq
	\label{equ211}
	\l[\til{\chi}_{lm_i}\l({\tilde{\xi},\til \t}\r),\til{\Pi}_{l'm'_i}\l({\tilde{\xi'},\til \t}\r) \r]=i \d_{ll'}\d_{m_im'_i}\d\l(\tilde{\xi}-\tilde{\xi'}\r)
	\eeq
	Using relations (\ref{equ214}) and expanding the Hamiltonian up to 
	second order in $\e$, we get,
	\bea
	\label{equ215}
	H&\simeq\dis\frac{1}{2}\sum_{_{l,m_i}}\dis\int_{\dis\til \t}^\infty d\til{r}\l[\pi^2_{lm_i}+\til{r}^D \dis\frac{\l(1-\e H_{_1}-\e^2 H_{_2}\r)^D}{\l(\dis 1+\e H_{_3}-\e^2 H_{_4}\r)^{1/2}}\l[\pa_{\til r} \frac{\sigma_{lm_i}}{\til r^{D/2}\l(1-\e H_{_1}-\e^2 H_{_2}\r)^{D/2}\l(\dis 1+\e H_{_3}-\e^2 H_{_4}\r)^{1/4}}\r]^2\r.\nn\\
	&\l.\dis+\frac{l(l+D-1)}{\til r^2\l(1-\e H_{_1}-\e^2 H_{_2}\r)^2 }\sigma^2_{lm_i}\r]
	\eea

	The Hamiltonian in Eq. (\ref{equ215})is of the form 
	\beq
	\label{Hamilt_11}
	H\simeq H_{_0}+ \e V_{_1}+\e^2 V_{_2} 
	\eeq
	where $ H_{_0}$ is the unperturbed scalar field Hamiltonian in the flat space-time, $V_{_1} \mbox{and}\,V_{_2}$ are the perturbed parts of the Hamiltonian given by;
	
	\br
	H_{_0}&=&\dis\frac{1}{2}\sum_{_{l,m_i}}\dis\int_{\dis\til \t}^\infty d\til{r}\l[\pi^2_{lm_i}+ \til r^D\dis\l[\pa_{\til r}\dis\frac{\sigma_{lm_i}}{\til r^{D/2}} \r]^2 +\frac{l(l+D-1)}{\til r^2}\sigma^2_{lm_i}\r]\\
	V_{_1}&=&\dis\frac{1}{2}\sum_{_{l,m_i}}\dis\int_{\dis\til \t}^\infty d\til{r}\l[\frac{\left(-D \,\sigma_{lm_i}+2\til r\, \sigma'_{lm_i}\right) \left(D  H_3 \,\sigma_{lm_i}-2 \til r H_3\,\sigma'_{lm_i}+2 D \til r H_1'\,\sigma_{lm_i} -\til r H_3'\, \sigma_{lm_i} \right)}{4 \til r^2} \r.\nn\\
	& &\l. \dis+\frac{2 l (l+D-1)  H_{_1}}{\til r^2}\sigma^2_{lm_i}\r]\\
	V_{_2}&=&\dis\frac{1}{2}\sum_{_{l,m_i}}\dis\int_{\dis\til \t}^\infty d\til{r}\l[\left(H^2_3+H_4\right)\,\sigma'^2_{lm_i}+
	\left(-\frac{D H^2_3}{\til r}-\frac{D H_4}{\til r}+D H_1 H_1'-D H_3 H_1'+D
	H_2'+H_3 H_3'
	+\frac{1}{2} H_4'\right)\,\sigma'_{lm_i}\sigma_{lm_i}\r.\nn\\
	&&\l.+\left(\frac{l (l+D-1) \left(3 H^2_1+2 H_2\right)}{\til r^2}+\frac{D^2 H_3^2}{4
		\til r^2}+\frac{D^2 H_4}{4 \til r^2}-\frac{D^2 H_1 H_1'}{2 \til r}+\frac{D^2 H_3 H_1'}{2 \til r}+\frac{1}{4} D^2 H_1'^2-\frac{D^2 H_2'}{2 \til r}\r.\r.\nn\\
	&&\l.\l.-\frac{D
		H_3H_3'}{2 \til r}-\frac{1}{4} D H_1' H_3'+\frac{1}{16} H_3'^2-\frac{d H_4'}{4 \til r}\right)\,\sigma^2_{lm_i}\r]
	\er
	where 
	\br
	H_{_1}= \dis\frac{\sqrt{1-f}}{\til r},  \quad  H_{_2}=\frac{1}{4\til r} \frac{\pa f}{\pa\til r},\quad
	H_{_3}=\dis\frac{1}{\sqrt{1-f}}\frac{\pa f}{\pa\til r}, \quad H_{_4}=\frac{-1}{4(1-f)}\l(\frac{\pa f}{\pa\til r}\r)^2+\frac{1}{2}\frac{\pa^2f}{\pa\til r^2}
	\er
	%
	and the redefined field operators are 
	\beq
	\til{\Pi}_{lm_i}=\frac{\pi_{lm_i}}{\l(1-f\r)^{1/4}} \quad\mbox{and}\quad \til\chi_{lm_i}=\frac{\sigma_{lm_i}}{\l(1-f\r)^{1/4}}
	\eeq
	such that they satisfy the following canonical commutation relation
	\beq
	\label{equ216}
	\l[\pi_{lm_i}(\til r, \til\t),\sigma_{l'm'_i}(\til r',\til \t)\r]=i \d_{ll'}\d_{m_im'_i}\d(\til r- \til r') 
	\eeq
	The Hamiltonian $H$ in Eq. (\ref{Hamilt_11}) is mapped to a system of $N$ coupled  time independent harmonic oscillators (HO) 
	with non-periodic boundary conditions. The interaction matrix elements of the Hamiltonian can be found in the Ref.\cite{dropbox}. 
	The total internal energy (E) and the entanglement entropy ($S_\a$) for the ground state of the HO's is computed numerically 
	as a function of $\e$  by using central difference scheme. 
	
	\section{Central Difference discretization }
	\label{app2}
	This is  one of the effective method for finding the approximate value for derivative of a function in the neighbourhood of any discrete point, $x_i=x_0+i\;h$,with unit steps of $h$. 
	The Taylor expansion of the function about the point  $x_0$ in the forward and backward difference scheme is given respectively by,
	\br
	f(x+h)= f(x) + \frac{ h f'(x) }{1!} +\frac{h^2 f''(x)}{2!}+ ......\\
	f(x-h)=f(x) - \frac{ h f'(x) }{1!} +\frac{h^2 f''(x)}{2!}- .......
	\er
which implies,
	\br
	f'(x)&= &\frac{f(x+h)-f(x-h)}{2 h}+ O(h^2)\\
	f''(x)&=&\frac{f(x+h)-2 f(x)+f(x-h)}{h^2}+O(h^2)\\
	f'''(x)&=&\frac{f(x+2 h)-2 f(x+h)+2 f(x-h)+f(x-2 h)}{2 h^3}+O(h^2)
	\er

	\section{ Plots of internal energy and EE as a function of $\e$ for different black hole space-times} 
	\label{app3}
	In this section of Appendix, we give plots of EE for different black hole  space-times;
	
	\begin{figure}[H]
		\centering
		\includegraphics[scale=2.1]{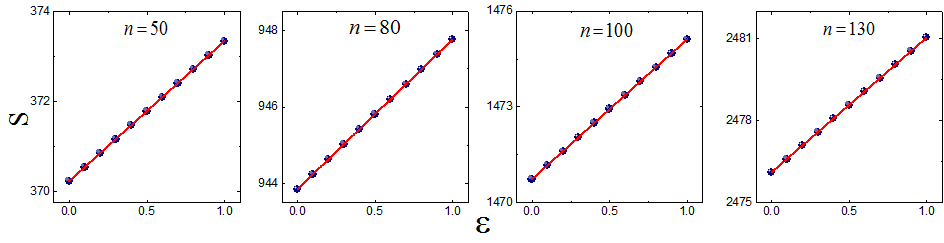}
		\caption{ Plots of the EE  as function of $\e$ for the 4-d Schwarzschild black hole with $N=300$, $n=50,80,100$, and $130$, respectively.  The blue dots are the numerical data and the red line is the best linear fit to the data.  }
		\label{fig2}
	\end{figure}
	
	\begin{figure}[H]
		\centering
		\includegraphics[scale=2.1]{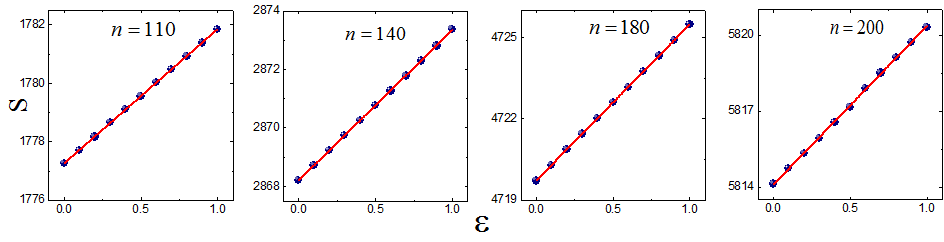}
		\caption{ \footnotesize {Plots of the EE  as function of $\e$ for the 4-d Schwarzschild black hole with $N=400$, $n=110,140,180$, and $200$ respectively.  The blue dots are the numerical data and the red line is the best linear fit to the data.  } }
		\label{fig16}
	\end{figure}
	
	\begin{figure}[H]
		\centering
		\includegraphics[scale=2.1]{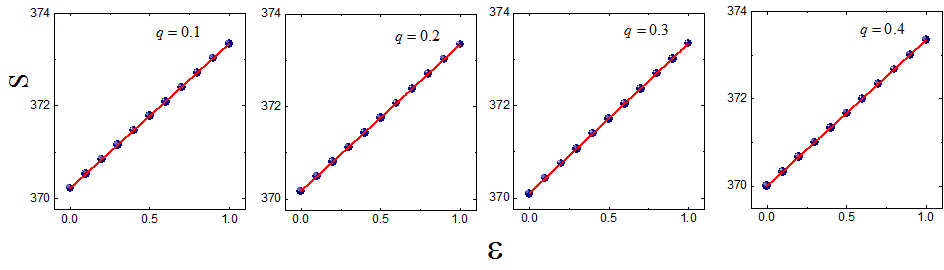}
		\caption{ \footnotesize {Plots of the EE of 4-d  R-N black hole in terms of $\e$ for different $q$'s with $N=300$, and  $n=50$. The blue dots are the numerical data and the red line is the best linear fit to the data.  } }
		\label{fig4}
	\end{figure}
\end{appendix}
		\end{widetext}


%
%
%


\newcommand{\noopsort}[1]{} \newcommand{\printfirst}[2]{#1}
\newcommand{\singleletter}[1]{#1} \newcommand{\switchargs}[2]{#2#1}
\providecommand{\href}[2]{#2}\begingroup\raggedright\endgroup

\end{document}